\documentclass[a4paper,12pt]{article}
\usepackage{amssymb}
\usepackage{amsmath}
\usepackage{amstext}

\newtheorem{theorem}{Theorem}
\newtheorem{lemma}[theorem]{Lemma}

\newtheorem{cor}[theorem]{Corollary}
\newtheorem{definition}[theorem]{Definition}
\newtheorem{rem}[theorem]{Remark}

\input{tcilatex}
\begin{document}

\title{On the Dynamics of solitons in the nonlinear Schroedinger equation}
\author{Vieri Benci\thanks{
Dipartimento di Matematica Applicata, Universit\`a degli Studi di Pisa, Via
F. Buonarroti 1/c, Pisa, ITALY and Department of Mathematics, College of Science, 
King Saud University, 
Riyadh, 11451, SAUDI ARABIA.
 e-mail: \texttt{benci@dma.unipi.it}} \and Marco Ghimenti\thanks{%
Dipartimento di Matematica Applicata, Universit\`a degli Studi di Pisa, Via
F. Buonarroti 1/c, Pisa, ITALY. 
e-mail: \texttt{ marco.ghimenti@dma.unipi.it, 
a.micheletti@dma.unipi.it}} \and Anna Maria Micheletti%
\addtocounter{footnote}{-1}\footnotemark }
\maketitle

\begin{abstract}
We study the behavior of the soliton solutions of the equation 
\begin{equation*}
i\frac{\partial \psi }{\partial t}=-\frac{1}{2m}\Delta \psi +\frac{1}{2}%
W_{\varepsilon }^{\prime }\left( \psi \right) +V(x)\psi
\end{equation*}%
where $W_{\varepsilon }^{\prime }$ is a suitable nonlinear term which is
singular for $\varepsilon =0.$ We use the \textquotedblleft
strong\textquotedblright\ nonlinearity to obtain results on existence,
shape, stability and dynamics of the soliton . The main result of this paper
(Theorem \ref{teo1}) shows that for $\varepsilon \rightarrow 0$ the orbit of
our soliton approaches the orbit of a classical particle in a potential $%
V(x) $.

\medskip


\medskip

\end{abstract}

\tableofcontents

\section{Introduction}

Roughly speaking a \textit{solitary wave} is a solution of a field equation
whose energy travels as a localized packet and which preserves this
localization in time. A \textit{soliton} is a solitary wave which exhibits
some strong form of stability so that it has a particle-like behavior. In
this paper we study the dynamics of solitons arising in the nonlinear
Schroedinger equation (NSE): 
\begin{equation}
i\frac{\partial \psi }{\partial t}=-\frac{1}{2m}\Delta \psi +\frac{1}{2}%
W_{\varepsilon }^{\prime }\left( \psi \right) +V(x)\psi .  \label{1}
\end{equation}%
A solution of our equation can be written as follows 
\begin{equation}
\psi (t,x)=\Psi _{\varepsilon }\left( t,x\right) +\varphi (t,x)  \label{uno}
\end{equation}%
where $\varphi (t,x)$ can be considered as a wave and $\Psi _{\varepsilon
}\left( t,x\right) $ is our soliton: a bump of energy concentrated in a ball
centered at the point $q=q_{\varepsilon }(t)$ with radius $R_{\varepsilon
}\rightarrow 0$ (for $\varepsilon \rightarrow 0$). Considering this
decomposition, the solutions to our equation can be thought as a combination
of a wave and a particle. $\varepsilon $ occurs in the equation as a
parameter. The main purpose of this paper is to show that for $\varepsilon $
and $\varphi (0,x)$ sufficiently small, our soliton behaves as a classical
particle in a potential $V(x)$. More exactly, we prove that the
decomposition (\ref{uno}) holds for all times and the bump follows a
dynamics which approaches the dynamics of a pointwise particle moving under
the action of the potential $V$ (Theorem \ref{teo1}); in particular the
position $q_{\varepsilon }(t)$ of the soliton approaches the position of the
particle uniformly on bounded time intervals (Corollary \ref{cor1}).

The attention of the mathematical community on the dynamics of soliton of
NSE began with the pioneering paper of Bronski and Jerrard \cite{BJ00}; then
Fr{\"{o}}hlich, Gustafson, Jonsson, and Sigal faced this problem using a
different approach \cite{FGJS06}. In the last years, several others works
appeared following the first approach (\cite{kera02}, \cite{kera06}, \cite%
{selv}, \cite{sq-ser}, \cite{squa}) or the second one (\cite{A08}, \cite{A09}%
, \cite{AFS09}, \cite{BP10}, \cite{GW08}, \cite{HZ08}, \cite{S09}).

In this paper, we have studied equation (\ref{1}) which gives a different
problem with respect to the ones mentioned above. Actually, in equation (\ref%
{1}), unlikely the other papers, the parameter $\varepsilon $ appears in the
nonlinear term and it will be chosen in such a way that $\left\vert \psi
\right\vert ^{2}$ approaches the delta-measure as $\varepsilon \rightarrow
0.\ $Thus equation (\ref{1}) describes the dynamics of a soliton when its
support is small with respect to the other relevant elements (namely $V(x)$,
the initial conditions and its $L^{2}$-size). Also the method employed here
is different from those of the paper quoted above and it exploits and
develops some ideas of \cite{BGM09}. Basically, we use the \textquotedblleft
strong\textquotedblright\ nonlinearity to obtain results on existence,
shape, stability and dynamics of the soliton (Theorem \ref{teo1}). Finally,
we notice that this method applies to a large class of nonlinearities and we
do not make any assumption on the nondegenaracy or the uniqueness of the
ground state solution (see the discussion in Section \ref{stp}).

\subsection{Notations\label{no}}

In the next we will use the following notations:

\begin{eqnarray*}
&&\func{Re}(z),\func{Im}(z)\text{ are the real and the imaginary part of }z
\\
&&B_{\rho }(x_{0})=B(x_{0},\rho )=\{x\in \mathbb{R}^{N}\ :\ |x-x_{0}|\leq
\rho \} \\
&&B(x_{0},\rho )^{C}=\mathbb{R}^{N}\smallsetminus B(x_{0},\rho ) \\
&&S_{\sigma }=\{u\in H^{1}\ :\ ||u||_{L^{2}}=\sigma \} \\
&&J(u) =\int \left( \frac{1}{2}\left\vert \nabla u\right\vert
^{2}+W(u)\right) dx \\
&&J_{\ }^{c}=\{u\in H^{1}\ :\ J_{\ }(u)<c\} \\
&&|\partial ^{\alpha }V(x)|=\sup_{i_{1},\dots ,i_{\alpha }}\left\vert \frac{%
\partial ^{\alpha }V(x)}{\partial x_{i_{1}}\dots \partial x_{i_{\alpha }}}%
\right\vert \text{ where }\alpha \in \mathbb{N},\ i_{1},\dots ,i_{\alpha
}\in \{1,\dots ,N\} \\
&&I_{\sigma ^{2}}=\inf_{u\in H^{1},\ \int u^{2}=\sigma ^{2}}J(u)=c \\
&&|\cdot |\text{ is the euclidean norm both of a vector or of a matrix} \\
&&\Gamma =\left\{ U\in H^{1},\ J(U)=\inf_{||U||_{L^{2}}=1}J(u)\right\} \text{
is the set of ground state solutions}
\end{eqnarray*}

\bigskip

\subsection{Statement of the problem\label{stp}}

First, we focus on the \textquotedblleft concentration\textquotedblright\
properties of a soliton solution of eq. (\ref{1}) without the potential term 
$V$. We consider the following Cauchy problem relative to the NSE: 
\begin{equation}
i\frac{\partial \psi }{\partial t}=-\frac{1}{2m}\Delta \psi +\frac{1}{2}%
W_{\varepsilon }^{\prime }\left( \psi \right)  \label{ch}
\end{equation}%
\begin{equation}
\psi \left( 0,x\right) =U_{\varepsilon }\left( x-\bar{q}\right) e^{i\bar{p}%
\cdot x}  \label{id}
\end{equation}%
where, with some abuse of notation, we have set%
\begin{equation}
W_{\varepsilon }\left( \psi \right) =\frac{1}{\varepsilon ^{N+2}}W\left(
\varepsilon ^{N/2}\left\vert \psi \right\vert \right) ;\ \ \ W_{\varepsilon
}^{\prime }\left( \psi \right) =\frac{1}{\varepsilon ^{N/2+2}}W^{\prime
}\left( \varepsilon ^{N/2}\left\vert \psi \right\vert \right) \frac{\psi }{%
\left\vert \psi \right\vert },  \label{ww}
\end{equation}%
and$\ W:\mathbb{R}^{+}\rightarrow \mathbb{R}$ is a real function which
satisfies suitable assumptions (see (ii) below). $U$ denotes a \emph{ground
state solution }of the equation%
\begin{equation}
-\frac{1}{2m}\Delta U+\frac{1}{2}W^{\prime }(U)=\omega _{1}U  \label{eq}
\end{equation}%
namely a function such that 
\begin{equation*}
\int \left( \frac{|\nabla U|^{2}}{2m}+W(U)\right) dx=c_{0}
\end{equation*}%
with 
\begin{equation}
c_{0}=\inf_{\substack{ ||u||_{L^{2}}=1  \\ u\in H^{1}}}\int \left( \frac{%
|\nabla u|^{2}}{2m}+W(u)\right) dx  \label{c0}
\end{equation}%
It is well known that we can choose $U$ radially symmetric and positive.

Direct computations show that, by virtue of (\ref{ww}), the function 
\begin{equation*}
U_{\varepsilon }(x)=\frac{1}{\varepsilon ^{N/2}}U\left( \frac{x}{\varepsilon 
}\right) .
\end{equation*}%
satisfies\ the equation%
\begin{equation}
-\frac{1}{2m}\Delta U_{\varepsilon }+\frac{1}{2}W_{\varepsilon }^{\prime
}(U_{\varepsilon })=\omega _{\varepsilon }U_{\varepsilon }.  \label{angela}
\end{equation}%
where 
\begin{equation*}
\omega _{\varepsilon }=\frac{\omega _{1}}{\varepsilon ^{2}}.
\end{equation*}%
Moreover $U_{\varepsilon }(x)$ is a ground state solution of (\ref{angela}).
In many cases, the ground state solution $U$ is unique up to translations
and change of sign, but we do not need this assumption.

Notice that the choice of $W_{\varepsilon }$ given by (\ref{ww}) implies
that 
\begin{equation*}
\left\Vert U_{\varepsilon }\right\Vert _{L^{2}}=1
\end{equation*}%
for every $\varepsilon >0$.

Direct computation shows that the solution of (\ref{ch}),(\ref{id}) is given
by the following soliton 
\begin{equation}
\psi _{q,\varepsilon }\left( t,x\right) =U_{\varepsilon }\left( x-\bar{q}-%
\bar{v}t\right) e^{i\left( \bar{p}\cdot x-\omega t\right) }\ \ \text{where\
\ }\ \bar{v}=\frac{\bar{p}}{m}\   \label{soli}
\end{equation}%
with 
\begin{equation*}
\omega =\omega _{\varepsilon }+\frac{1}{2}m\bar{v}^{2}
\end{equation*}

Thus $\psi _{q,\varepsilon }\left( t,x\right) $ behaves as a particle of
\textquotedblleft radius\textquotedblright\ $\varepsilon $ living in the
point 
\begin{equation}
q=\bar{q}+\bar{v}t  \label{vu}
\end{equation}%
Since $\left\Vert \psi _{q,\varepsilon }\left( t,\cdot \right) \right\Vert
_{L^{2}}=1$ for every $\varepsilon >0,$ if $\varepsilon \rightarrow 0,$ we
have that 
\begin{equation*}
\left\vert \psi _{q,\varepsilon }\left( t,x\right) \right\vert
^{2}\rightarrow \delta \left( x-\bar{q}-\bar{v}t\right) \ \ \text{in}\ \ 
\mathcal{D}^{\prime }\left( \mathbb{R}^{N}\right) \ \ \forall t\in \mathbb{R}%
,
\end{equation*}%
where $\delta \left( x-x_{0}\right) $ denotes the Dirac measure concentrated
in the point $x_{0}.$ The energy $E_{\varepsilon }\left( \psi \right) $ of
the configuration $\psi $ is given by 
\begin{equation*}
E_{\varepsilon }(\psi )=\int \left[ \frac{1}{2m}\left\vert \nabla \psi
\right\vert ^{2}+W_{\varepsilon }(\psi )\right] dx,
\end{equation*}%
so the energy of $\psi _{q,\varepsilon }\ $is 
\begin{equation}
E_{\varepsilon }\left( \psi _{q,\varepsilon }\right) =\int \left( \frac{%
|\nabla U_{\varepsilon }|^{2}}{2m}+W_{\varepsilon }(U_{\varepsilon })\right)
dx+\frac{1}{2}m\bar{v}^{2}  \label{ciccia}
\end{equation}

Thus $\psi _{q,\varepsilon }\left( t,x\right) $ behaves as a particle of
mass $m$: $\bar{p}$ can be interpreted as its momentum, $\frac{1}{2m}\bar{p}%
^{2}=\frac{1}{2}m\bar{v}^{2}$ as its kinetic energy and%
\begin{equation*}
\int \frac{|\nabla U_{\varepsilon }|^{2}}{2m}+W_{\varepsilon
}(U_{\varepsilon })\ dx=\frac{c_{0}}{\varepsilon ^{2}}
\end{equation*}%
as the \textit{internal} energy; here $c_{0}$ is a constant defined as
follows%
\begin{equation*}
c_{0}:=\int \frac{|\nabla U_{1}|^{2}}{2m}+W(U_{1})\ dx
\end{equation*}

The aim of this paper is to study the dynamics of the solitons in the
presence of a potential $V(x)$ namely to investigate the problem

\begin{equation}
\left\{ 
\begin{array}{l}
i\frac{\partial \psi }{\partial t}=-\frac{1}{2m}\Delta \psi +\frac{1}{2}%
W_{\varepsilon }^{\prime }(\psi )+V(x)\psi \\ 
\\ 
\psi \left( 0,x\right) =\psi _{0,\varepsilon }(x)%
\end{array}%
\right.  \tag{$P$}  \label{schr}
\end{equation}%
where $\psi _{0,\varepsilon }$ satisfies the following assumptions%
\begin{eqnarray}
\psi _{0,\varepsilon }(x) &=&U_{\varepsilon }\left( x-\bar{q}\right) e^{i%
\bar{p}\cdot x}+\varphi _{0,\varepsilon }(x),\ \ \ \varphi _{0,\varepsilon
}\in H^{1}(\mathbb{R}^{N});  \label{phih} \\
\left\Vert \psi _{0,\varepsilon }\right\Vert _{L^{2}} &=&1  \label{p2} \\
E_{\varepsilon }\left( \psi _{0,\varepsilon }\right) &\leq &\frac{c_{0}}{%
\varepsilon ^{2}}+M  \label{p3}
\end{eqnarray}%
with $M>0$ independent of $\varepsilon ;\ $here $E_{\varepsilon }\left( \psi
\right) $ denotes the energy in the presence of the potential $V:$%
\begin{equation*}
E_{\varepsilon }(\psi )=\int \left[ \frac{1}{2m}\left\vert \nabla \psi
\right\vert ^{2}+W_{\varepsilon }(\psi )+V(x)\left\vert \psi \right\vert ^{2}%
\right] dx.
\end{equation*}

\subsection{The main results}

We make the following assumptions:

\begin{description}
\item (i) $W:\mathbb{R}^{+}\rightarrow \mathbb{R}$ is a $C^{2}$ function
which satisfies the following assumptions: 
\begin{equation}
W(0)=W^{\prime }(0)=W^{\prime \prime }(0)=0  \tag{$W_{0}$}  \label{W}
\end{equation}%
\begin{equation}
|W^{\prime \prime }(s)|\leq c_{1}|s|^{q-2}+c_{2}|s|^{p-2}\text{ for some }%
2<q\leq p<2^{\ast }  \tag{$W_{1}$}  \label{Wp}
\end{equation}%
\begin{equation}
W(s)\geq -c|s|^{\nu },\text{ }c\geq 0,\ 2<\nu <2+\frac{4}{N}\text{ for }s%
\text{ large}  \tag{$W_{2}$}  \label{W0}
\end{equation}%
\begin{equation}
\exists s_{0}\in \mathbb{R}^{+}\text{ such that }W(s_{0})<0  \tag{$W_{3}$}
\label{W1}
\end{equation}

\item (ii) $V:\mathbb{R}^{N}\rightarrow \mathbb{R}$ is a $C^{2}$-function
with bounded derivatives which satisfies the following assumptions: 
\begin{equation}
0\leq V(x)\leq V_{0}<\infty ;  \tag{$V_{0}$}  \label{V0}
\end{equation}
\end{description}

The main result of this paper is the following theorem which describes the
shape and the dynamics of the soliton $\Psi _{\varepsilon }\left( t,x\right)$%
:

\begin{theorem}
\label{teo1} Assume (i) and (ii); then the solution of problem (\ref{schr})
has the following form 
\begin{equation}
\psi _{\varepsilon }(t,x)=\Psi _{\varepsilon }\left( t,x\right) +\varphi
_{\varepsilon }(t,x)  \label{formosa2}
\end{equation}%
where $\Psi _{\varepsilon }\left( t,x\right) $ is a function having support
in a ball $B_{R_{\varepsilon }}(q_{\varepsilon }),$ with radius $%
R_{\varepsilon }\rightarrow 0$ and center $q_{\varepsilon }=q_{\varepsilon
}(t)$. Moreover,%
\begin{equation}
\left\Vert \left\vert \Psi _{\varepsilon }\left( t,x\right) \right\vert
-U_{\varepsilon }\left( x-q_{\varepsilon }(t)\right) \right\Vert
_{L^{2}}\rightarrow 0\ \ as\;\;\varepsilon \rightarrow 0  \label{conc}
\end{equation}%
uniformly in $t,$ where $U_{\varepsilon }=\frac{1}{\varepsilon ^{N/2}}%
U\left( \frac{x}{\varepsilon }\right) $ and $U$ is a ground state solution
of (\ref{eq}).

The dynamics is given by the following equations: 
\begin{equation}
\left\{ 
\begin{array}{l}
\dot{q}_{\varepsilon }(t)=\frac{1}{m_{\varepsilon }(t)}\ p_{\varepsilon
}(t)+K_{\varepsilon }(t) \\ 
\\ 
\dot{p}_{\varepsilon }(t)=-\nabla V(q_{\varepsilon }(t))+F_{\varepsilon
}(q_{\varepsilon })+H_{\varepsilon }(t)%
\end{array}%
\right.  \label{probcauchy}
\end{equation}%
with initial data 
\begin{equation}
\left\{ 
\begin{array}{l}
q_{\varepsilon }(0)=\bar{q}+o(1) \\ 
{p}_{\varepsilon }(0)=\bar{p}+o(1)%
\end{array}%
\right.
\end{equation}%
where

\begin{itemize}
\item (a) $q_{\varepsilon }(t)$ is the \textbf{barycenter}$\ $of the soliton
and it has the following\ form:%
\begin{equation*}
q_{\varepsilon }(t)=\frac{\displaystyle\int x\ \left\vert \Psi _{\varepsilon
}\right\vert ^{2}dx}{\displaystyle\int \left\vert \Psi _{\varepsilon
}\right\vert ^{2}dx}
\end{equation*}

\item (b) $\displaystyle m_{\varepsilon }(t)=m\int \left\vert \Psi
_{\varepsilon }\left( t,x\right) \right\vert ^{2}dx=m+o(1)$ can be
interpreted as the \textbf{mass} of the soliton,

\item (c)$\ p_{\varepsilon }(t)$ is the \textbf{momentum}$\ $of the soliton
and it has the following\ form: 
\begin{equation*}
p_{\varepsilon }(t)=\func{Im}\int \nabla \Psi _{\varepsilon }\left(
t,x\right) \ \overline{\Psi _{\varepsilon }\left( t,x\right) }\ dx;
\end{equation*}

\item (d) $K_{\varepsilon }(t)\ $and $H_{\varepsilon }(t)$\ are errors due
to the fact that the soliton is not a point and 
\begin{equation*}
\sup_{t\in \mathbb{R}}\left( |H_{\varepsilon }(t)|+|K_{\varepsilon
}(t)|\right) \rightarrow 0\ \ \ \text{as}\;\;\varepsilon \rightarrow 0.
\end{equation*}

\item (e)\ $F_{\varepsilon }(q_{\varepsilon })$ is the force due to the
pressure of the wave $\varphi _{\varepsilon }$ on the soliton and $%
F_{\varepsilon }\rightarrow 0\ $in the space of distributions, more exactly
we have that 
\begin{equation*}
\forall \tau _{0},\tau _{1},\ \left\vert \int_{\tau _{0}}^{\tau
_{1}}F_{\varepsilon }(q_{\varepsilon })\ dt\right\vert \leq c(\varepsilon
)\left( 1+\left\vert \tau _{1}-\tau _{0}\right\vert \right) 
\end{equation*}%
where $c(\varepsilon )\rightarrow 0$ as $\varepsilon \rightarrow 0.$
\end{itemize}
\end{theorem}

\begin{cor}
\label{cor1}Let\ $\mathfrak{q}$ and $\mathfrak{p}$ be the solution of the
following Cauchy problem:%
\begin{equation}
\left\{ 
\begin{array}{l}
\mathfrak{\dot{q}}(t)=\frac{1}{m}\ \mathfrak{p}(t) \\ 
\\ 
\mathfrak{\dot{p}}(t)=-\nabla V(\mathfrak{q}(t))%
\end{array}%
\right.  \label{got}
\end{equation}%
with initial data 
\begin{equation}
\left\{ 
\begin{array}{l}
\mathfrak{q}(0)=q_{\varepsilon }(0) \\ 
\mathfrak{p}(0)={p}_{\varepsilon }(0)%
\end{array}%
\right.  \label{got1}
\end{equation}%
where $q_{\varepsilon }(t)$ and ${p}_{\varepsilon }(t)$ are as in Th. \ref%
{teo1}. Then, as $\varepsilon \rightarrow 0$ 
\begin{equation*}
\left( q_{\varepsilon }(t),{p}_{\varepsilon }(t)\right) \rightarrow \left( 
\mathfrak{q}(t),\mathfrak{p}(t)\right)
\end{equation*}%
uniformly on compact sets.
\end{cor}

Let us discuss the set of our assumptions.

\begin{rem}
The conditions (\ref{W}) and (\ref{V0}) are assumed for simplicity; in fact
they can be weakened as follows 
\begin{equation*}
W(0)=W^{\prime }(0)=0,\ \ W^{\prime \prime }(0)=E_{0}
\end{equation*}%
and 
\begin{equation*}
E_{1}\leq V(x)\leq V_{\infty }<+\infty .
\end{equation*}%
In fact, in the general case, the solution of the Schroedinger equation is
modified only by a phase factor.
\end{rem}

\begin{rem}
\label{remgss} In \cite{BBGM07} the authors prove that if (ii) holds
equation (\ref{ch}) admits orbitally stable solitary waves having the form (%
\ref{id}). In particular the authors show that, under assumptions (\ref{W}),
(\ref{Wp}), (\ref{W0}) and (\ref{W1}), for any $\sigma $ there exists a
minimizer $U(x)=U_{\sigma }(x)$ of the functional 
\begin{equation*}
J(u)=\int \left( \frac{1}{2}\left\vert \nabla u\right\vert ^{2}+W(u)\right)
dx
\end{equation*}%
on the manifold $S_{\sigma }:=\{u\in H^{1},\ ||u||_{L^{2}}=\sigma \}$. Such
a minimizer satisfies eq.(\ref{eq}) where $\omega _{1}$ is a Lagrange
multiplier.
\end{rem}

\begin{rem}
\label{kato}By our assumptions, the problem (\ref{schr}) has a unique
solution 
\begin{equation}
\psi \in C^{0}(\mathbb{R},H^{2}(\mathbb{R}^{N}))\cap C^{1}(\mathbb{R},L^{2}(%
\mathbb{R}^{N})).  \label{gv}
\end{equation}%
Let us recall a result on the global existence of solutions of the Cauchy
problem (\ref{schr}) (see \cite{Ca03,GV79,Ka89}). Assume (\ref{Wp}), (\ref%
{W0}) and (\ref{W1}) for $W$. Let $D(A)$ (resp. $D(A^{1/2})$) denote the
domain of the self-adjoint operator $A$ (resp. $A^{1/2}$) where 
\begin{equation*}
A=-\Delta +V:L^{2}(\mathbb{R}^{N})\rightarrow L^{2}(\mathbb{R}^{N}).
\end{equation*}%
If $V\geq 0$, $V\in C^{2}$ and $|\partial ^{2}V|\in L^{\infty }$ and the
initial data $\psi (0,x)\in D(A^{1/2})$ then there exists the global
solution $\psi $ of (\ref{schr}) and

\begin{equation*}
\psi (t,x)\in C^{0}\left( \mathbb{R},D(A^{1/2})\right) \cap C^{1}(\mathbb{R}%
,H^{-1}(\mathbb{R}^{N})).
\end{equation*}%
Furthermore, if $\psi (0,x)\in D(A)$ then 
\begin{equation*}
\psi (t,x)\in C^{0}(\mathbb{R},D(A))\cap C^{1}(\mathbb{R},L^{2}(\mathbb{R}%
^{N})).
\end{equation*}%
In this case, since $D(A)\subset H^{2}(\mathbb{R}^{N}),$ (\ref{gv}) is
satisfied.
\end{rem}

\section{The dynamics of solitons\label{TDS}}

In this section, after having stated the main dynamical properties of our
system (Subsection \ref{fio}), we define explicitely the splitting (\ref{uno}%
) and the equations which rules the dynamics of the soliton. In all this
section, it is not necessary to assume $\varepsilon $ to be small.

\subsection{First integrals of NSE\label{fio}}

Equation (\ref{schr}) is the Euler-Lagrange equation relative to the
Lagrangian density 
\begin{equation}
\mathcal{L}=\func{Re}(i\partial _{t}\psi \bar{\psi})-\frac{1}{2m}\left\vert
\nabla \psi \right\vert ^{2}-W_{\varepsilon }(\psi )-V(x)\left\vert \psi
\right\vert ^{2}.  \label{lagr}
\end{equation}%
Sometimes it is useful to write $\psi $ in polar form 
\begin{equation}
\psi (t,x)=u(t,x)e^{iS(t,x)}.  \label{polar1}
\end{equation}%
Thus the state of the system ${\psi }$ is uniquely defined by the couple of
variables $(u,S)\in \mathbb{R}^{+}\times \left[ \mathbb{R}/\left( 2\pi 
\mathbb{Z}\right) \right] $. Using these variables, the action $\mathcal{S=}%
\int \mathcal{L}dxdt$ takes the form 
\begin{equation}
\mathcal{S}(u,S)=-\int \left[ \frac{1}{2m}\left\vert \nabla u\right\vert
^{2}+W_{\varepsilon }(u)+\left( \partial _{t}S+\frac{1}{2m}\left\vert \nabla
S\right\vert ^{2}+V(x)\right) u^{2}\right] dxdt  \label{polarAZ}
\end{equation}%
and equation (\ref{schr}) splits in the two following equations:

\begin{equation}
-\frac{1}{2m}\Delta u+W_{\varepsilon }(u)+\left( \partial _{t}S+\frac{1}{2m}%
\left\vert \nabla S\right\vert ^{2}+V(x)\right) u=0  \label{Sh1}
\end{equation}

\begin{equation}
\partial _{t}\left( u^{2}\right) +\nabla \cdot \left( u^{2}\frac{\nabla S}{m}%
\right) =0  \label{Sh2}
\end{equation}

Noether's theorem states that any invariance for a one-parameter group of
the Lagrangian implies the existence of an integral of motion (see e.g. \cite%
{Gelfand} or \cite{milano}). They are derived by a continuity equation.

Now we describe the first integrals which will be relevant for this paper,
namely the energy, the ``hylenic charge'' and the momentum.

\bigskip

\textbf{Energy.}\emph{\ }The energy, by definition, is the quantity which is
preserved by the time invariance of the Lagrangian; it has the following
form 
\begin{equation}
E_{\varepsilon }(\psi )=\int \left[ \frac{1}{2m}\left\vert \nabla \psi
\right\vert ^{2}+W_{\varepsilon }(\psi )+V(x)\left\vert \psi \right\vert ^{2}%
\right] dx.
\end{equation}%
Using (\ref{polar1}) we get: 
\begin{equation}
E_{\varepsilon }(\psi )=\int \left( \frac{1}{2m}\left\vert \nabla
u\right\vert ^{2}+W_{\varepsilon }(u)\right) dx+\int \left( \frac{1}{2m}%
\left\vert \nabla S\right\vert ^{2}+V(x)\right) u^{2}dx  \label{Schenergy}
\end{equation}%
Thus the energy has two components: the \textit{internal energy} (which,
sometimes, is also called \textit{binding energy}) 
\begin{equation}
J_{\varepsilon }(u)=\int \left( \frac{1}{2m}\left\vert \nabla u\right\vert
^{2}+W_{\varepsilon }(u)\right) dx  \label{j}
\end{equation}%
and the \textit{dynamical energy} 
\begin{equation}
G(u,S)=\int \left( \frac{1}{2m}\left\vert \nabla S\right\vert
^{2}+V(x)\right) u^{2}dx  \label{g}
\end{equation}%
which is composed by the \textit{kinetic energy} $\frac{1}{2m}\int
\left\vert \nabla S\right\vert ^{2}u^{2}dx$ and the \textit{potential energy}
$\int V(x)u^{2}dx$. By our assumptions, the internal energy is bounded from
below and the dynamical energy is positive.

\bigskip

\textbf{Hylenic charge. }Following \cite{milano} the \textit{hylenic charge}%
, is defined as the quantity which is preserved by the invariance of the
Lagrangian with respect to the action 
\begin{equation*}
\psi \mapsto e^{i\theta }\psi .
\end{equation*}%
This invariance gives the continuity equation (\ref{Sh2}). Thus, in this
case, the charge is nothing else but the $L^{2}$ norm, namely: 
\begin{equation*}
C(\psi )=\int \left\vert \psi \right\vert ^{2}dx=\int u^{2}dx.
\end{equation*}

\bigskip

\textbf{Momentum. }The momentum is constant in time if the Lagrangian is
space-translation invariant; this happens when $V$ is a constant. In general
we have the equation 
\begin{equation}
\partial _{t}\left( u^{2}\nabla S\right) =-u^{2}\nabla V+\nabla \cdot T
\label{20}
\end{equation}%
where $T$ is the \textit{stress tensor} whose components are given by 
\begin{eqnarray}
T_{jk} &=&\func{Re}\left( \partial _{x_{j}}\psi \partial _{x_{k}}\bar{\psi}%
\right)  \notag \\
&&-\delta _{jk}\left[ \frac{1}{4m}\Delta |\psi |^{2}-\frac{1}{2}%
W_{\varepsilon }^{\prime }\left( \psi \right) \bar{\psi}+W_{\varepsilon
}(\psi )\right]  \label{stress} \\
&=&\left[ \partial _{x_{j}}u\ \partial _{x_{k}}u+\left( \partial _{x_{j}}S\
\partial _{x_{k}}S\right) u^{2}-\frac{1}{4m}\delta _{jk}\Delta \left(
u^{2}\right) \right]  \notag \\
&&+\delta _{jk}\left( \frac{1}{2}W_{\varepsilon }^{\prime
}(u)u-W_{\varepsilon }(u)\right)  \label{stress2}
\end{eqnarray}

\textbf{Proof:} It is well known that the stress tensor has the following
form (see e.g. \cite{milano} or \cite{sammomme}) 
\begin{equation*}
T_{jk}=-\func{Re}\left( \frac{\partial \mathcal{L}}{\partial \left( \partial
_{x_{j}}\psi \right) }\cdot \partial _{x_{k}}\bar{\psi}\right) +\delta _{jk}%
\mathcal{L}
\end{equation*}%
Using equation (\ref{schr}) to eliminate $i\partial _{t}\psi ,$ we get 
\begin{eqnarray*}
\mathcal{L} &=&\func{Re}(i\partial _{t}\psi \bar{\psi})-\frac{1}{2m}%
\left\vert \nabla \psi \right\vert ^{2}-W_{\varepsilon }(\psi
)-V(x)\left\vert \psi \right\vert ^{2} \\
&=&\func{Re}\left[ \left( -\frac{1}{2m}\Delta \psi +\frac{1}{2}%
W_{\varepsilon }^{\prime }\left( \psi \right) +V(x)\psi \right) \bar{\psi}%
\right] -\frac{1}{2m}\left\vert \nabla \psi \right\vert ^{2}-W_{\varepsilon
}(\psi )-V(x)\left\vert \psi \right\vert ^{2} \\
&=&\func{Re}\left( -\frac{1}{2m}\Delta \psi \bar{\psi}\right) -\frac{1}{2m}%
\left\vert \nabla \psi \right\vert ^{2}+\frac{1}{2}W_{\varepsilon }^{\prime
}\left( \psi \right) \bar{\psi}-W_{\varepsilon }(\psi ).
\end{eqnarray*}

Moreover, we have that%
\begin{eqnarray*}
\Delta \psi \bar{\psi} &=&\Delta \psi _{1}\psi _{1}+\Delta \psi _{2}\psi
_{2}=\nabla \cdot \left( \psi _{1}\nabla \psi _{1}\right) +\nabla \cdot
\left( \psi _{2}\nabla \psi _{2}\right) -\left\vert \nabla \psi \right\vert
^{2} \\
&=&\frac{1}{2}\Delta |\psi _{1}|^{2}+\frac{1}{2}\Delta |\psi
_{2}|^{2}-\left\vert \nabla \psi \right\vert ^{2}=\frac{1}{2}\Delta |\psi
|^{2}-\left\vert \nabla \psi \right\vert ^{2}
\end{eqnarray*}%
Then%
\begin{equation*}
\mathcal{L}=-\frac{1}{4m}\Delta |\psi |^{2}+\frac{1}{2}W_{\varepsilon
}^{\prime }\left( \psi \right) \bar{\psi}-W_{\varepsilon }(\psi )
\end{equation*}

We recall that, if $z=a+ib$ is a complex number, by definition, $\frac{%
\partial L}{\partial z}=\frac{\partial L}{\partial a}+i\frac{\partial L}{%
\partial b}$. So 
\begin{equation*}
\func{Re}\left( \frac{\partial \mathcal{L}}{\partial \left( \partial
_{x_{j}}\psi \right) }\cdot \partial _{x_{k}}\bar{\psi}\right) =-\func{Re}%
\left( \partial _{x_{j}}\psi \partial _{x_{k}}\bar{\psi}\right)
\end{equation*}%
and the conclusion follows from direct computation.

$\square $

If $V=const$, eq. (\ref{20}) is a continuity equation and the momentum 
\begin{equation}
\mathbf{P}(\psi )=\int u^{2}\nabla S\ dx=\int \func{Im}\left( \bar{\psi}%
\nabla \psi \right) dx  \label{P}
\end{equation}%
is constant in time. Notice that, by equation (\ref{Sh2}), the density of
momentum $u^{2}\nabla S$ is nothing else than the flow of hylenic charge.

Let us consider the soliton (\ref{soli}); in this case we have%
\begin{equation*}
E_{\varepsilon }\left( \psi _{q,\varepsilon }\right) =\frac{c_{0}^{2}}{%
\varepsilon ^{2}}+\frac{1}{2m}\bar{p}^{2}
\end{equation*}%
where $c_{0}$ is defined by (\ref{c0}),%
\begin{equation*}
C(\psi _{\varepsilon })=1
\end{equation*}%
and%
\begin{equation*}
\mathbf{P}(\psi _{\varepsilon })=\bar{p}.
\end{equation*}

Now, let us see the rescaling properties of the internal energy and the $%
L^{2}$ norm of a function $u(x)$ having the form 
\begin{equation*}
u(x):=\frac{1}{\varepsilon ^{N/2}}v\left( \frac{x}{\varepsilon }\right).
\end{equation*}%
We have 
\begin{equation*}
||u||_{L^{2}}^{2}=\frac{1}{\varepsilon ^{N}}\int v\left( \frac{x}{%
\varepsilon }\right) ^{2}dx=\int v\left( \xi \right) ^{2}d\xi
=||v||_{L^{2}}^{2}.
\end{equation*}%
and

\begin{eqnarray*}
J_{\varepsilon }(u) &=&\int \left[ \frac{1}{2m}|\nabla u|^{2}+W_{\varepsilon
}(u)\right] dx \\
&=&\int \left[ \frac{1}{2m}|\nabla u|^{2}+\frac{1}{\varepsilon ^{N+2}}%
W(\varepsilon ^{N/2}u)\right] dx \\
&=&\int \left[ \frac{1}{2m}\frac{1}{\varepsilon ^{N}}\left\vert \nabla
_{x}v\left( \frac{x}{\varepsilon }\right) \right\vert ^{2}+\frac{1}{%
\varepsilon ^{N+2}}W\left( v\left( \frac{x}{\varepsilon }\right) \right) %
\right] dx \\
&=&\int \left[ \frac{1}{2m}\frac{1}{\varepsilon ^{N+2}}\left\vert \nabla
_{\xi }v\left( \xi \right) \right\vert ^{2}+\frac{1}{\varepsilon ^{N+2}}%
W\left( v\left( \xi \right) \right) \right] \varepsilon ^{N}d\xi \\
&=&\frac{1}{\varepsilon ^{2}}\int \frac{1}{2m}\left\vert \nabla _{\xi
}v\left( \xi \right) \right\vert ^{2}+W\left( v\left( \xi \right) \right)
d\xi =\frac{1}{\varepsilon ^{2}}J_{1}(u).
\end{eqnarray*}

\subsection{Definition of the soliton\label{DDSS}}

In this section we want describe a method to split a solution of eq. (\ref%
{schr}) in a wave and a soliton as in (\ref{uno}).

If our solution has the following form 
\begin{equation*}
\psi _{\varepsilon }(t,x)=u_{\varepsilon }e^{iS_{\varepsilon }}=\left[
U_{\varepsilon }(x-\xi (t))+w_{\varepsilon }(t,x)\right] e^{iS_{\varepsilon
}(t,x)}
\end{equation*}%
where $w$ is sufficiently small, then a possible choice is to identify the
soliton with $U(x-\xi (t))e^{iS(t,x)}$ and the wave with $w_{\varepsilon
}(t,x)e^{iS(t,x)}.$ However, we want to give a definition which localize the
soliton, namely to assume the function $\Psi _{\varepsilon }\left(
t,x\right) $ to have compact support in space.

Roughly speaking, the soliton can be defined as the part of the field $\psi
_{\varepsilon }$ where some density function $\rho _{\varepsilon }(t,x)$ is
sufficiently large (e.g, after a suitable normalization, $\rho _{\varepsilon
}(t,x)\geq 1$).

For the moment we do not define $\rho _{\varepsilon }(t,x)$ explicitely. We
just require that $\rho _{\varepsilon }(t,x)$ satisfies the follwing
assumptions:

\begin{itemize}
\item $\rho _{\varepsilon }\in C^{1}(\mathbb{R}^{N+1})\ $and $\rho
_{\varepsilon }(t,x)\rightarrow 0$ as $|x|\ \rightarrow \infty $

\item $\rho _{\varepsilon }$ satisfies the continuity equation%
\begin{equation}
\partial _{t}\rho _{\varepsilon }+\nabla \cdot J_{\rho _{\varepsilon }}=0
\label{a2}
\end{equation}%
for some $J_{\rho _{\varepsilon }}\in C^{1}(\mathbb{R}^{N+1})$
\end{itemize}

In order to fix the ideas you may think of $\rho _{\varepsilon }(t,x)$ as a
smooth approximation of $u_{\varepsilon }(t,x)^{2}.$ An explicit definition
of $\rho _{\varepsilon }(t,x)$ is given in Section \ref{dd}. However, in
other problems, it might be more useful to make different choices of it such
as the energy density. We have postponed the choice of $\rho _{\varepsilon }$
since the results of this section are independent of this choice.

Next we set%
\begin{equation*}
\chi _{\varepsilon }(t,x)=\sqrt{\varphi \left( \rho _{\varepsilon
}(t,x)\right) }
\end{equation*}%
where 
\begin{equation*}
\varphi \left( s\right) =\left\{ 
\begin{array}{cl}
0 & \text{ if } s\leq 1 \\ 
s-1 & \text{ if } 1\leq s\leq 2 \\ 
1 & \text{ if } s\geq 2%
\end{array}%
\right.
\end{equation*}

So we have that $\chi _{\varepsilon }(t,x)=1$ where $\rho _{\varepsilon
}(t,x)\geq 2$ and $\chi _{\varepsilon }(t,x)=0$ where $\rho _{\varepsilon
}(t,x)\leq 1:$ thus you may think of $\chi _{\varepsilon }(t,x)$ as a sort
of approximation of the characteristic function of the region occupied by
the soliton. Finally, we set 
\begin{eqnarray}
\Psi _{\varepsilon }\left( t,x\right) &=&\psi _{\varepsilon }(t,x)\chi
_{\varepsilon }  \label{aa} \\
\varphi _{\varepsilon }(t,x) &=&\psi _{\varepsilon }(t,x)\left[ 1-\chi
_{\varepsilon }\right]  \label{bb}
\end{eqnarray}

\ $\Psi _{\varepsilon }\left( t,x\right) $ is the soliton and $\varphi
_{\varepsilon }(t,x)\ $is the wave; the region 
\begin{eqnarray}
\Sigma _{\varepsilon ,t} &=&\left\{ (t,x)\in \mathbb{R}^{N+1}|\ 1<\rho
(t,x)<2\right\}  \label{giovanna} \\
&=&\left\{ (t,x)\in \mathbb{R}^{N+1}|\ 0<\chi _{\varepsilon }(t,x)<1\right\}
\end{eqnarray}%
is the region where the soliton and the wave interact with each other; we
will refer to it as the \textit{halo }of the soliton.

\bigskip

\subsection{The equation of dynamics of the soliton}

\begin{definition}
\label{DS}We define the following quantities relative to the soliton:

\begin{itemize}
\item the \emph{barycenter}: 
\begin{equation*}
q_{\varepsilon }(t)=\frac{\int x\ \left\vert \Psi _{\varepsilon }\right\vert
^{2}dx}{\int \left\vert \Psi _{\varepsilon }\right\vert ^{2}dx}
\end{equation*}

\item the \emph{momentum}: 
\begin{equation*}
p_{\varepsilon }(t)=\int \nabla S_{\varepsilon }\left\vert \Psi
_{\varepsilon }\right\vert ^{2}dx
\end{equation*}

\item the \emph{mass:} 
\begin{equation*}
m_{\varepsilon }(t)=m\int \left\vert \Psi _{\varepsilon }\right\vert ^{2}dx.
\end{equation*}
\end{itemize}
\end{definition}

\begin{rem}
Notice that the mass of the soliton $m_{\varepsilon }(t)$ depends on $t.$
The global mass is constant (namely $m$) but it is shared between the
soliton and the wave whose mass is $\int u_{\varepsilon }^{2}\left[ 1-\chi
_{\varepsilon }^{2}\right] dx$
\end{rem}

The next theorem shows the relation between $q_{\varepsilon }(t)$ and $%
p_{\varepsilon }(t)$ and their derivatives.

\begin{theorem}
\textbf{\label{kappa2} }The following equations hold 
\begin{eqnarray}
\dot{q}_{\varepsilon } &=&\frac{p_{\varepsilon }}{m_{\varepsilon }}+\frac{1}{%
m_{\varepsilon }}\int_{\Sigma _{\varepsilon ,t}}(x-q_{\varepsilon })\left[
u_{\varepsilon }^{2}\nabla S_{\varepsilon }\cdot \nabla \rho _{\varepsilon
}-\nabla \cdot J_{\rho _{\varepsilon }}\right] dx  \label{qpunto3} \\
&&  \notag \\
\dot{p}_{\varepsilon } &=&-\int \nabla V\left\vert \Psi _{\varepsilon
}\right\vert ^{2}dx-\int_{\Sigma _{\varepsilon ,t}}\left[ T\cdot \nabla \rho
_{\varepsilon }+u_{\varepsilon }^{2}\nabla S_{\varepsilon }\left( \nabla
\cdot J_{\rho _{\varepsilon }}\right) \right] dx.  \label{ppunto2}
\end{eqnarray}
\end{theorem}

\begin{rem}
The term $\int_{\Sigma _{\varepsilon ,t}}T\cdot \nabla \rho _{\varepsilon
}dx $ represents the pressure of the wave on the soliton; if $\varepsilon
\rightarrow 0\ $and $\partial \Sigma _{\varepsilon ,t}$ is sufficiently
regular then 
\begin{equation*}
\int_{\Sigma _{\varepsilon ,t}}T\cdot \nabla \rho _{\varepsilon
}dx=\int_{\sigma _{\varepsilon }}T\cdot \mathbf{n}\ d\sigma
\end{equation*}%
where $\sigma _{\varepsilon }=\left\{ x\ |\ \rho _{\varepsilon
}(x)=1\right\} $ and $\mathbf{n}$ is its outer normal.
\end{rem}

\noindent \textbf{Proof of Th. \ref{kappa2}.} We calculate the first
derivative of the barycenter.

\begin{eqnarray*}
\dot{q}_{\varepsilon }(t) &=&\frac{d}{dt}\left( \frac{\int x\left\vert \Psi
_{\varepsilon }\right\vert ^{2}dx}{\int \left\vert \Psi _{\varepsilon
}\right\vert ^{2}dx}\right) \\
&=&\frac{\frac{d}{dt}\int x\left\vert \Psi _{\varepsilon }\right\vert ^{2}dx%
}{\int \left\vert \Psi _{\varepsilon }\right\vert ^{2}dx}-\frac{\left( \int
x\left\vert \Psi _{\varepsilon }\right\vert ^{2}dx\right) \left( \frac{d}{dt}%
\int \left\vert \Psi _{\varepsilon }\right\vert ^{2}dx\right) }{\left( \int
\left\vert \Psi _{\varepsilon }\right\vert ^{2}dx\right) ^{2}} \\
&=&\frac{\frac{d}{dt}\int x\left\vert \Psi _{\varepsilon }\right\vert ^{2}dx%
}{\int \left\vert \Psi _{\varepsilon }\right\vert ^{2}dx}-q_{\varepsilon }(t)%
\frac{\frac{d}{dt}\int \left\vert \Psi _{\varepsilon }\right\vert ^{2}dx}{%
\int \left\vert \Psi _{\varepsilon }\right\vert ^{2}dx}=\frac{\int
(x-q_{\varepsilon }(t))\frac{d}{dt}(\left\vert \Psi _{\varepsilon
}\right\vert ^{2})dx}{\int \left\vert \Psi _{\varepsilon }\right\vert ^{2}dx}%
.
\end{eqnarray*}%
We have%
\begin{equation*}
\nabla \chi ^{2}=\nabla \varphi \left( \rho _{\varepsilon }(t,x)\right)
=\varphi ^{\prime }\left( \rho _{\varepsilon }(t,x)\right) \nabla \rho
_{\varepsilon }=\mathbb{I}_{\Sigma _{\varepsilon ,t}}\nabla \rho
_{\varepsilon }
\end{equation*}%
and 
\begin{eqnarray*}
\frac{d}{dt}\chi ^{2} &=&\frac{d}{dt}\varphi \left( \rho _{\varepsilon
}(t,x)\right) =\varphi ^{\prime }\left( \rho _{\varepsilon }(t,x)\right)
\partial _{t}\rho _{\varepsilon } \\
&& \\
&=&\mathbb{I}_{\Sigma _{\varepsilon ,t}}\partial _{t}\rho _{\varepsilon }=-%
\mathbb{I}_{\Sigma _{\varepsilon ,t}}\nabla \cdot J_{\rho _{\varepsilon }}
\end{eqnarray*}%
where $\mathbb{I}_{\Sigma _{\varepsilon ,t}}$ is the characteristic function
of $\Sigma _{\varepsilon ,t}.$

So, we have

\begin{eqnarray}
\dot{q}_{\varepsilon }(t) &=&\frac{\int (x-q_{\varepsilon }(t))\frac{d}{dt}%
(\chi ^{2}u_{\varepsilon }^{2})dx}{\int \left\vert \Psi _{\varepsilon
}\right\vert ^{2}dx}  \notag \\
&=&\frac{\int (x-q_{\varepsilon }(t))\left( \chi ^{2}\frac{d}{dt}%
u_{\varepsilon }^{2}+u_{\varepsilon }^{2}\frac{d}{dt}\chi ^{2}\right) dx}{%
\int \left\vert \Psi _{\varepsilon }\right\vert ^{2}dx}  \label{qpunto} \\
&=&\frac{\int_{\mathbb{R}^{N}}(x-q_{\varepsilon }(t))\chi ^{2}\frac{d}{dt}%
u_{\varepsilon }^{2}dx-\int_{\Sigma _{\varepsilon ,t}}(x-q_{\varepsilon
}(t))\nabla \cdot J_{\rho _{\varepsilon }}dx}{\int_{\mathbb{R}%
^{N}}\left\vert \Psi _{\varepsilon }\right\vert ^{2}dx}.  \notag
\end{eqnarray}

For the first term we use the continuity equation (\ref{Sh2}). We have

\begin{eqnarray*}
\int (x-q_{\varepsilon }(t))\chi ^{2}\frac{d}{dt}u_{\varepsilon }^{2}dx
&=&\int (x-q_{\varepsilon }(t))\nabla \cdot \left( u_{\varepsilon }^{2}\frac{%
\nabla S_{\varepsilon }}{m}\right) \ \chi ^{2}dx \\
&=&\frac{1}{m}\int \left( u_{\varepsilon }^{2}\nabla S_{\varepsilon }\right)
\ \chi ^{2}dx+\frac{1}{m}\int (x-q_{\varepsilon }(t))u_{\varepsilon
}^{2}\nabla S_{\varepsilon }\cdot \nabla \chi ^{2}dx \\
&=&\frac{p_{\varepsilon }(t)}{m}+\frac{1}{m}\int_{\Sigma _{\varepsilon
,t}}(x-q_{\varepsilon }(t))u_{\varepsilon }^{2}\nabla S_{\varepsilon }\cdot
\nabla \rho _{\varepsilon }dx.
\end{eqnarray*}%
Concluding, we get the first equation of motion: 
\begin{equation*}
\dot{q}_{\varepsilon }(t)=\frac{p_{\varepsilon }(t)}{m_{\varepsilon }}+\frac{%
\int_{\Sigma _{\varepsilon ,t}}(x-q_{\varepsilon }(t))\left[ u_{\varepsilon
}^{2}\nabla S_{\varepsilon }\cdot \nabla \rho _{\varepsilon }-\nabla \cdot
J_{\rho _{\varepsilon }}\right] dx}{m_{\varepsilon }}
\end{equation*}

Next, we will get the second one. We have that 
\begin{equation}
\dot{p}_{\varepsilon }=\int \left( \frac{\partial }{\partial t}%
u_{\varepsilon }^{2}\nabla S_{\varepsilon }\right) \chi ^{2}dx+\int
u_{\varepsilon }^{2}\nabla S_{\varepsilon }\frac{\partial }{\partial t}\chi
^{2}dx.  \label{lina}
\end{equation}

Now, using (\ref{20}) we have that 
\begin{eqnarray*}
\int \left( \frac{\partial }{\partial t}u_{\varepsilon }^{2}\nabla
S_{\varepsilon }\right) \chi ^{2}dx &=&-\int \nabla V\left\vert \Psi
_{\varepsilon }\right\vert ^{2}dx+\int \nabla \cdot T\ \chi ^{2}dx \\
&=&-\int \nabla V\left\vert \Psi _{\varepsilon }\right\vert ^{2}dx-\int
T\cdot \nabla \chi ^{2}dx \\
&=&-\int \nabla V\left\vert \Psi _{\varepsilon }\right\vert
^{2}dx-\int_{\Sigma _{\varepsilon ,t}}T\cdot \nabla \rho _{\varepsilon }dx.
\end{eqnarray*}%
The second term of eq. (\ref{lina}) takes the form: 
\begin{equation*}
\int u_{\varepsilon }^{2}\nabla S_{\varepsilon }\frac{\partial }{\partial t}%
\chi ^{2}dx=-\int_{\Sigma _{\varepsilon ,t}}u_{\varepsilon }^{2}\nabla
S_{\varepsilon }\left( \nabla \cdot J_{\rho _{\varepsilon }}\right) dx.
\end{equation*}

$\square $\bigskip

It is possible to give a \textquotedblleft pictorial\textquotedblright\
interpretation to equations (\ref{qpunto3}) and (\ref{ppunto2}). We may
assume that $u_{\varepsilon }^{2}$ represents the density of a fluid;$\ $so
the soliton is a bump of fluid particles which stick together and the halo $%
\Sigma _{\varepsilon ,t}$ can be regarded as the interface where the soliton
and the wave might exchange particles, momentum and energy.

Hence,

\begin{itemize}
\item $m_{\varepsilon }(t)$ is the mass of the soliton

\item $\frac{\nabla S_{\varepsilon }}{m}$ is the velocity of the fluid
particles and $\nabla S_{\varepsilon }$ is their momentum
\end{itemize}

So each term of the equations (\ref{qpunto3}) and (\ref{ppunto2}) have the
following interpretation

\begin{itemize}
\item $\frac{p_{\varepsilon }(t)}{m_{\varepsilon }}$ is the average velocity
of each particle; in fact 
\begin{equation*}
\frac{p_{\varepsilon }(t)}{m_{\varepsilon }}=\frac{\int \nabla
S_{\varepsilon }\left\vert \Psi _{\varepsilon }\right\vert ^{2}dx}{%
m_{\varepsilon }}=\frac{\int \frac{\nabla S_{\varepsilon }}{m}\left\vert
\Psi _{\varepsilon }\right\vert ^{2}dx}{\int \left\vert \Psi _{\varepsilon
}\right\vert ^{2}dx}
\end{equation*}

\item the ``halo term'' $\frac{1}{m_{\varepsilon }}\int_{\Sigma
_{\varepsilon ,t}}(x-q_{\varepsilon })\left[ u_{\varepsilon }^{2}\nabla
S_{\varepsilon }\cdot \nabla \rho _{\varepsilon }-\nabla \cdot J_{\rho
_{\varepsilon }}\right] dx$ describes the change of the average velocity of
the soliton due to the exchange of fluid particles

\item the term $-\int \nabla V\left\vert \Psi _{\varepsilon }\right\vert
^{2}dx$ describes the volume force acting on the soliton

\item the term $-\int_{\Sigma _{\varepsilon ,t}}T\cdot \nabla \rho
_{\varepsilon }dx$ describes the surface force exerted by the wave on the
soliton

\item the term$\ -\int_{\Sigma _{\varepsilon ,t}}u_{\varepsilon }^{2}\nabla
S_{\varepsilon }\left( \nabla \cdot J_{\rho _{\varepsilon }}\right) dx\ $%
describes the change of the momentum of the soliton due to the exchange of
fluid particles with the wave.
\end{itemize}

\section{The limit dynamics}

In this section, we analyze the dynamics of the soliton as $\varepsilon
\rightarrow 0$ and we end proving the main theorem i.e. Th. \ref{teo1}.

\subsection{Analysis of the concentration point of the soliton}

If $\psi _{\varepsilon }(t,x)$ is a solution of the problem (\ref{schr}), we
say that $\hat{q}_{\varepsilon }(t)$ is the \textit{concentration point} of $%
\psi _{\varepsilon }(t,x)$ if it minimizes the following quantity 
\begin{equation}
f(q)=\left\Vert |\psi _{\varepsilon }(t,x)|-U_{\varepsilon }(x-q)\right\Vert
_{L^{2}}^{2}.  \label{maria}
\end{equation}

It is easy to see that $f(q)$ has a minimizer; of course, it might happen
that it is not unique; in this case we denote by $\hat{q}_{\varepsilon }(t)$
one of the minimizers of $f$ at the time $t.$

Basically $\hat{q}_{\varepsilon }(t)$ is a good candidate for the position
of our soliton, but it cannot satisfy an equation of type (\ref{probcauchy})
since in general it is not uniquely defined and \textit{a fortiori} is not
differentiable. $\hat{q}_{\varepsilon }(t)$ could be uniquely defined if we
make assumptions on the non degeneracy of the ground state, but we do not
like to make such assumptions since they are very hard to be verified and in
general they do not hold. Actually the position of the soliton is supposed
to be $q_{\varepsilon }(t)$ as in Def. \ref{DS}. However, as we will see, $%
\hat{q}_{\varepsilon }$ is useful to recover some estimates on $%
q_{\varepsilon }.$ So, in this subsection we will analyze some properties of 
$\hat{q}_{\varepsilon }(t).$ We start with a variant of a result contained
in \cite{BBGM07}.

\begin{lemma}
\label{proto} Given $u\in H^{1}$, we define (if it exists) $\hat{q}\in 
\mathbb{R}^{N}$ to be a minimizer of the function 
\begin{equation*}
q\mapsto ||U(x-q)-u(x)||_{L^{2}}^{2}.
\end{equation*}%
For any $\eta $ there exists a $\delta (\eta )$ such that, if $u\in
J^{c_{0}+\delta (\eta )}\cap S_{1}$ (see section \ref{no}), $\hat{q}$ exists
and it holds 
\begin{equation}
||U(x-\hat{q})-u||_{H^{1}}\leq \eta  \label{eq:conc1}
\end{equation}%
\begin{equation}
\int\limits_{\mathbb{R}^{N}\smallsetminus B(\hat{q},\hat{R}_{\eta
})}u^{2}dx\leq \eta  \label{eq:conc1bis}
\end{equation}%
where $\hat{R}_{\eta }=-C\log (\eta )$ and $U\in \Gamma $.
\end{lemma}

\textbf{Proof}: The proof of (\ref{eq:conc1}) can be found in \cite{BBGM07}.
If $U\in\Gamma$, again by \cite{BBGM07} we know that, for $R$ sufficiently
large, 
\begin{equation*}
\int_{|x|>R}U^{2}(x)dx<\int_{|x|>R}C_{1}e^{-C_{2}|x|}.
\end{equation*}

Thus 
\begin{equation*}
\int_{|x|>R}U^{2}(x)dx=C_{3}\int_{R}^{\infty }\rho ^{N-1}e^{-C_{2}\rho
}d\rho =C_{4}R^{N}e^{-C_{2}R}\leq e^{-C_{5}R}
\end{equation*}
where the $C_{i}$'s are suitable positive constants. We remark that $R$ does
not depend on $U$.

Now, it is sufficient to take $\hat{R}_{\eta }>-\frac{1}{C_{5}}\log (\eta )$
and by (\ref{eq:conc1}) we obtain (\ref{eq:conc1bis}).

$\square $

\bigskip

We define the set of admissible initial data as follows:

\begin{equation*}
B_{\varepsilon ,M}=\left\{ \psi (x)=U_{\varepsilon }\left( x-q_{0}\right)
e^{ip_{0}\cdot x}+\varphi (x):E_{\varepsilon }\left( \psi \right) \leq \frac{%
c_{0}}{\varepsilon ^{2}}+M\text{ and }\left\Vert \psi \right\Vert
_{L^{2}}=1\right\}
\end{equation*}

\bigskip

\begin{lemma}
\label{protor}For every $\eta >0,$ there exists $\varepsilon =\varepsilon
(\eta )>0$ such that 
\begin{equation}
\int\limits_{\mathbb{R}^{N}\smallsetminus B(\hat{q}_{\varepsilon
},\varepsilon \hat{R}_{\eta })}\left\vert \psi _{\varepsilon
}(t,x)\right\vert ^{2}dx<\eta  \label{brutta}
\end{equation}%
where $\psi _{\varepsilon }(t,x)$ is a solution of problem (\ref{schr}),
with initial data in $B_{\varepsilon ,M}$ and $\hat{q}_{\varepsilon }$ is
the concentration point of $\psi _{\varepsilon }$.
\end{lemma}

\textbf{Proof.} By the conservation law, the energy $E_{\varepsilon }(\psi
_{\varepsilon }(t,x))$ is constant with respect to $t$. Then we have, by
hypothesis on the initial datum 
\begin{equation*}
E_{\varepsilon }(\psi _{\varepsilon }(t,x))=E_{\varepsilon }(\psi
_{\varepsilon }(0,x))\leq \frac{c_{0}}{\varepsilon ^{2}}+M.
\end{equation*}%
Thus 
\begin{eqnarray}
J_{\varepsilon }(\psi _{\varepsilon }(t,x)) &=&E_{\varepsilon }(\psi
(t,x))-G(\psi (t,x))  \notag  \label{Jh} \\
&=&E_{\varepsilon }(\psi _{\varepsilon }(t,x))-\int_{\mathbb{R}^{N}}\left[ 
\frac{|\nabla S_{\varepsilon }(t,x)|^{2}}{2m}+V(x)\right] u_{\varepsilon
}(t,x)^{2}dx  \notag \\
&\leq &E_{\varepsilon }(\psi _{\varepsilon }(t,x))\leq \frac{c_{0}}{%
\varepsilon ^{2}}+M
\end{eqnarray}%
because $V\geq 0$. By rescaling the inequality (\ref{Jh}), and setting $%
y=x/\varepsilon $ we get 
\begin{equation}
J(|\varepsilon ^{N/2}\psi _{\varepsilon }(t,\varepsilon y)|)\leq
c_{0}+\varepsilon ^{2}M
\end{equation}%
We choose $\varepsilon $ small such that $\varepsilon ^{2}M\leq \delta (\eta
).$ Then $\varepsilon ^{N/2}\psi _{\varepsilon }(t,\varepsilon y)\in
J^{c_{0}+\delta (\eta )}\cap S_{1},$ and so applying Lemma \ref{proto}. 
\begin{equation}
\int\limits_{\mathbb{R}^{N}\smallsetminus B(\hat{q},\hat{R}_{\eta
})}\varepsilon ^{N}|\psi _{\varepsilon }(t,\varepsilon y)|^{2}dy<\eta
\end{equation}%
Now, making the change of variable $x=\varepsilon y,$ we obtain the desired
result.

\bigskip $\square $

\begin{lemma}
\label{lemmaconc}If $\psi _{\varepsilon }(t,x)$ is a solution of problem (%
\ref{schr}), with initial data in $B_{\varepsilon ,M}$ and $\varepsilon $
sufficiently small, then 
\begin{equation}
\int\limits_{\mathbb{R}^{N}\smallsetminus B\left( \hat{q}_{\varepsilon },%
\sqrt{\varepsilon }\right) }\left\vert \psi _{\varepsilon }(t,x)\right\vert
^{2}dx=\eta (\varepsilon )  \label{epsn}
\end{equation}%
where $\eta (\varepsilon )\rightarrow 0$ as $\varepsilon \rightarrow 0$
\end{lemma}

\textbf{Proof. }First we prove that for every $\eta >0,$ there exists $%
\varepsilon _{1}\left( \eta \right) >0$ such that, if $\psi _{\varepsilon
}(0,x)\in B_{\varepsilon _{1}\left( \eta \right) ,M}$, we have 
\begin{equation*}
\int\limits_{\mathbb{R}^{N}\smallsetminus B\left( \hat{q}_{\varepsilon },%
\sqrt{\varepsilon _{1}\left( \eta \right) }\right) }\left\vert \psi
_{\varepsilon }(t,x)\right\vert ^{2}dx<\eta.
\end{equation*}%
Arguing as in the proof of Lemma \ref{protor}, if $\varepsilon _{1}\left(
\eta \right) \leq \min \left[ \sqrt{\frac{\delta (\eta )}{M}},\frac{1}{\hat{R%
}_{\eta }^{2}}\right] ,$ we get (\ref{brutta}). At this point, since $%
\varepsilon _{1}\left( \eta \right) \leq \frac{1}{\hat{R}_{\eta }^{2}}$ we
have that $\varepsilon _{1}\left( \eta \right) \hat{R}_{\eta }\leq \sqrt{%
\varepsilon _{1}\left( \eta \right) }$.

Now set 
\begin{equation*}
\varepsilon \left( \eta \right) =\min_{\eta \leq \zeta }
\varepsilon_{1}\left( \zeta \right).
\end{equation*}
Clearly, $\varepsilon \left( \eta \right) $ is a non-increasing function
(which might be discontinuous) and $\varepsilon \left( \eta
\right)\rightarrow 0$ as $\eta \rightarrow 0$. Then it has a
``pseudoinverse'' function $\eta (\varepsilon )$ namely a function which is
the inverse in the monotonicity points, which is discontinuous where $%
\varepsilon \left( \eta\right)$ is constant and constant where $\varepsilon
\left( \eta \right) $ is discontinuous. Moreover $\eta (\varepsilon )$ as $%
\varepsilon \rightarrow0$

$\square $

\subsection{Definition of the density $\protect\rho _{\protect\varepsilon }$%
\label{dd}}

First of all we notice that, in Lemma \ref{lemmaconc}, it is not restrictive
to assume that%
\begin{equation}
\eta =\eta (\varepsilon )\geq \varepsilon.  \label{laida}
\end{equation}%
Now we set%
\begin{equation*}
\rho _{\varepsilon }(t,x)=a(x)\ast u(t,x)^{2}
\end{equation*}%
where, $a_{\varepsilon }(s)\in C^{\infty },$%
\begin{equation*}
a_{\varepsilon }(s)=\left\{ 
\begin{array}{cc}
3 & |s|\leq \eta ^{\frac{1}{8}}\left( 1-\eta ^{\frac{1}{8}}\right) \\ 
0 & |s|\geq \eta ^{\frac{1}{8}}\left( 1+\eta ^{\frac{1}{8}}\right)%
\end{array}%
\right.
\end{equation*}%
and%
\begin{equation}
\left\vert \nabla a_{\varepsilon }(s)\right\vert \leq \eta ^{-\frac{1}{4}}.
\label{laida2}
\end{equation}

\begin{lemma}
\label{lemmino} Take $\psi _{\varepsilon }$ a solution of (\ref{schr}) with
initial data in $B_{\varepsilon ,M}$. 
\begin{equation*}
\begin{array}{cc}
\text{If }|x-\hat{q}_{\varepsilon }(t)|\ \leq \eta ^{\frac{1}{8}}\left(
1-2\eta ^{\frac{1}{8}}\right) & \ \text{then\ }\rho _{\varepsilon }(t,x)\geq
3\left( 1-\eta \right) \\ 
\text{if }|x-\hat{q}_{\varepsilon }(t)|\ \geq \eta ^{\frac{1}{8}}\left(
1+2\eta ^{\frac{1}{8}}\right) & \ \text{then\ }\rho _{\varepsilon }(t,x)\leq
3\eta%
\end{array}%
\end{equation*}%
where $\eta =\eta (\varepsilon )$ as in Lemma \ref{lemmaconc}.\ In
particular we have that 
\begin{equation}
\Sigma _{\varepsilon ,t}\subset B\left( \hat{q}_{\varepsilon }(t),\eta ^{%
\frac{1}{8}}\left( 1+2\eta ^{\frac{1}{8}}\right) \right) \setminus B\left( 
\hat{q}_{\varepsilon }(t),\eta ^{\frac{1}{8}}\left( 1-2\eta ^{\frac{1}{8}%
}\right) \right)  \label{gianna}
\end{equation}%
where $\Sigma _{\varepsilon ,t}$ is defined by (\ref{giovanna}).
\end{lemma}

\textbf{Proof.} If $|x-\hat{q}_{\varepsilon }|\ \leq \eta ^{\frac{1}{8}%
}\left( 1-2\eta ^{\frac{1}{8}}\right) ,$ then 
\begin{equation*}
|x-\hat{q}_{\varepsilon }|\ +\sqrt{\varepsilon }\leq \eta ^{\frac{1}{8}%
}\left( 1-2\eta ^{\frac{1}{8}}\right) +\sqrt{\eta }\leq \eta ^{\frac{1}{8}%
}\left( 1-\eta ^{\frac{1}{8}}\right)
\end{equation*}%
and hence%
\begin{equation*}
B(\hat{q}_{\varepsilon },\ \sqrt{\varepsilon })\subset B\left(x,\ \eta ^{%
\frac{1}{8}}\left( 1-\eta ^{\frac{1}{8}}\right) \right).
\end{equation*}

Then, by using Lemma \ref{lemmaconc},%
\begin{eqnarray*}
\rho _{\varepsilon }(t,x) &=&\int a_{\varepsilon }(y-x)u_{\varepsilon
}(t,y)^{2}dy\geq 3\int_{B(x,\ \eta ^{1/8}-\eta ^{1/4})}u_{\varepsilon
}(t,y)^{2}dy \\
&\geq &3\int_{B(\hat{q}_{\varepsilon },\ \varepsilon ^{1/2})}u_{\varepsilon
}(t,y)^{2}dy\geq 3\left( 1-\eta \right) .
\end{eqnarray*}

If $|x-\hat{q}_{\varepsilon }(t)|\ \geq \eta ^{\frac{1}{8}}\left( 1+2\eta ^{%
\frac{1}{8}}\right) ,$ 
\begin{equation*}
|x-\hat{q}_{\varepsilon }|\ -\sqrt{\varepsilon }\geq \eta ^{\frac{1}{8}%
}\left( 1+2\eta ^{\frac{1}{8}}\right) -\sqrt{\eta }\geq \eta ^{\frac{1}{8}%
}\left( 1+\eta ^{\frac{1}{8}}\right)
\end{equation*}%
and so%
\begin{equation*}
B\left( x,\ \eta ^{\frac{1}{8}}\left( 1+\eta ^{\frac{1}{8}}\right) \right)
\subset \mathbb{R}^{N}\smallsetminus B(\hat{q}_{\varepsilon },\ \sqrt{%
\varepsilon }).
\end{equation*}

Then, using again Lemma \ref{lemmaconc},%
\begin{eqnarray*}
\rho _{\varepsilon }(t,x) &=&\int a_{\varepsilon }(y-x)u_{\varepsilon
}(t,y)^{2}dy\leq 3\int_{B\left( x,\ \eta ^{\frac{1}{8}}\left( 1+\eta ^{\frac{%
1}{8}}\right) \right) }u_{\varepsilon }(t,y)^{2}dy \\
&\leq &3\int_{\mathbb{R}^{N}\smallsetminus B(\hat{q}_{\varepsilon },\ \sqrt{%
\varepsilon })}u_{\varepsilon }(t,y)^{2}dy\leq 3\eta
\end{eqnarray*}

$\square $

Clearly, $\rho _{\varepsilon }=a_{\varepsilon }\ast u_{\varepsilon }^{2}\in
C^{1}(\mathbb{R}^{N+1})$ and, by (\ref{Sh2}), it satisfies the continuity
equation (\ref{a2}) with 
\begin{equation}
J_{\rho _{\varepsilon }}=a_{\varepsilon }\ast \left( u_{\varepsilon
}^{2}\nabla S_{\varepsilon }\right) .  \label{bruttina}
\end{equation}

Therefore, the results of Section \ref{TDS} hold. In particular, we have
that the support of $\Psi _{\varepsilon }\left( t,x\right) $ is contained in 
$B\left( \hat{q}_{\varepsilon },\ \eta ^{\frac{1}{8}}\left( 1+2\eta ^{\frac{1%
}{8}}\right) \right) $ when $\eta $ is sufficiently small (namely $\eta <1/3$%
). Moreover, by (\ref{gianna}), we see that the size of the halo is an
infinitesimal of higher order with respect to the diameter of the soliton.

\subsection{The equation of dynamics as $\protect\varepsilon \rightarrow 0$}

\begin{theorem}
\textbf{\label{kappa} }The following equations hold 
\begin{equation}
\dot{q}_{\varepsilon }(t)=\frac{p_{\varepsilon }(t)}{m_{\varepsilon }(t)}%
+K_{\varepsilon }(t)  \label{qpunto2}
\end{equation}%
\begin{equation}
\dot{p}_{\varepsilon }=-\nabla V(q_{\varepsilon }(t))+F_{\varepsilon
}(q_{\varepsilon })+H_{\varepsilon }(t)  \label{ppunto3}
\end{equation}%
where 
\begin{equation}
\sup_{t\in \mathbb{R}}\left( |H_{\varepsilon }(t)|+|K_{\varepsilon
}(t)|\right) \rightarrow 0\ \ \text{as\ \ }\varepsilon \rightarrow 0
\label{qpuntostima}
\end{equation}%
and%
\begin{equation}
F_{\varepsilon }(q_{\varepsilon })=-\int_{\Sigma _{\varepsilon ,t}}T\cdot
\nabla \rho _{\varepsilon }dx.  \label{effe}
\end{equation}%
Moreover we have that 
\begin{equation}
\forall \tau _{0},\tau _{1},\ \left\vert \int_{\tau _{0}}^{\tau
_{1}}F_{\varepsilon }(q_{\varepsilon })\ dt\right\vert \leq c(\varepsilon
)\left( 1+\left\vert \tau _{1}-\tau _{0}\right\vert \right)  \label{finale}
\end{equation}%
where $c(\varepsilon )\rightarrow 0$ as $\varepsilon \rightarrow 0.$
\end{theorem}

\noindent \textbf{Proof.} We set 
\begin{equation*}
K_{\varepsilon }(t)=\frac{1}{m_{\varepsilon }}\int_{\Sigma _{\varepsilon
,t}}(x-q_{\varepsilon })\left[ u_{\varepsilon }^{2}\nabla S_{\varepsilon
}\cdot \nabla \rho _{\varepsilon }-\nabla \cdot J_{\rho _{\varepsilon }}%
\right] dx,
\end{equation*}%
\begin{equation*}
H_{1,\varepsilon }(t)=\int_{\Sigma _{\varepsilon ,t}}u_{\varepsilon
}^{2}\nabla S_{\varepsilon }\left( \nabla \cdot J_{\rho _{\varepsilon
}}\right) dx,
\end{equation*}%
\begin{equation*}
H_{2,\varepsilon }(t)=\nabla V(q_{\varepsilon }(t))-\int \nabla
V(x)\left\vert \Psi _{\varepsilon }\right\vert ^{2}dx,
\end{equation*}%
\begin{equation*}
H_{\varepsilon }(t)=H_{1,\varepsilon }(t)+H_{2,\varepsilon }(t),
\end{equation*}%
and hence, by Th. \ref{kappa2}, we need just to prove (\ref{qpuntostima}).

We estimate each individual term of $K_{\varepsilon }.$ We have that

\begin{equation}
\underset{x\in \Sigma _{\varepsilon ,t}}{\sup }\left\vert x-q_{\varepsilon
}\right\vert \leq 2\left( \eta ^{1/8}+2\eta ^{1/4}\right) \leq 3\eta ^{1/8}
\label{ghi1}
\end{equation}%
since $q_{\varepsilon }(t),x\in B(\hat{q}_{\varepsilon },\ \eta ^{1/8}+2\eta
^{1/4}).$

Also, by (\ref{laida2}) and well known properties on convolutions, 
\begin{eqnarray}
\underset{x\in \Sigma _{\varepsilon ,t}}{\sup }\left\vert \nabla \rho
_{\varepsilon }\right\vert &\leq &\ \underset{x\in \mathbb{R}^{N}}{\sup }%
\left\vert \nabla a_{\varepsilon }(x)\ast u_{\varepsilon
}(t,x)^{2}\right\vert  \label{ghi2} \\
&\leq &\left\Vert \nabla a_{\varepsilon }\right\Vert _{L^{\infty }}\cdot
\left\Vert u_{\varepsilon }\right\Vert _{L^{2}}^{2}\leq \frac{1}{\eta ^{1/4}}
\notag
\end{eqnarray}

If $\psi _{\varepsilon }(0,x)\in B_{\varepsilon ,M}$, by (\ref{g}), we have 
\begin{equation}
G(\psi )=E_{\varepsilon }(\psi )-J_{\varepsilon }(\psi )\leq \frac{c_{0}}{%
\varepsilon ^{2}}+M-\frac{c_{0}}{\varepsilon ^{2}}=M;  \label{stimaG}
\end{equation}%
so, by Lemma \ref{lemmaconc},%
\begin{eqnarray}
\int_{\mathbb{R}^{N}\smallsetminus B(\hat{q}_{\varepsilon },\ \sqrt{%
\varepsilon })}u^{2}|\nabla S_{\varepsilon }|\ &\leq &\left( \int_{\mathbb{R}%
^{N}\smallsetminus B(\hat{q}_{\varepsilon },\ \sqrt{\varepsilon }%
)}u_{\varepsilon }^{2}\right) ^{\frac{1}{2}}\left( \int_{\mathbb{R}%
^{N}\smallsetminus B(\hat{q}_{\varepsilon },\ \sqrt{\varepsilon }%
)}u_{\varepsilon }^{2}|\nabla S_{\varepsilon }|^{2}\right) ^{\frac{1}{2}} 
\notag \\
&\leq &\eta ^{\frac{1}{2}}\left[ 2mG(\psi )\right] ^{\frac{1}{2}}\leq
const.\eta ^{\frac{1}{2}}  \label{ghi2.5}
\end{eqnarray}%
and in particular

\begin{equation}
\int_{\Sigma _{\varepsilon ,t}}u^{2}|\nabla S_{\varepsilon }|\ dx\leq \eta ^{%
\frac{1}{2}}\left[ 2mG(\psi )\right] ^{\frac{1}{2}}\leq const.\eta ^{\frac{1%
}{2}}.  \label{ghi3}
\end{equation}

Finally, by (\ref{bruttina})%
\begin{eqnarray}
\underset{x\in \Sigma _{\varepsilon ,t}}{\sup }\left\vert \nabla \cdot
J_{\rho _{\varepsilon }}\right\vert &\leq &\ \underset{x\in \mathbb{R}^{N}}{%
\sup }|\left( \nabla \cdot a_{\varepsilon }\right) \ast \left(
u_{\varepsilon }^{2}\nabla S_{\varepsilon }\right) |  \notag \\
&\leq &\left\Vert \nabla \cdot a_{\varepsilon }\right\Vert _{L^{\infty
}}\cdot \int_{\mathbb{R}^{N}}u_{\varepsilon }^{2}|\nabla S_{\varepsilon }|\ 
\notag \\
&\leq &\frac{1}{\eta ^{1/4}}\left( \int_{\mathbb{R}^{N}}u_{\varepsilon
}^{2}\right) ^{\frac{1}{2}}\cdot \left( \int_{\mathbb{R}^{N}}u_{\varepsilon
}^{2}|\nabla S_{\varepsilon }|\right) ^{\frac{1}{2}}  \notag \\
&\leq &\frac{\left[ 2mG(\psi )\right] ^{\frac{1}{2}}}{\eta ^{1/4}}%
=const.\eta ^{-\frac{1}{4}}.  \label{ghi4}
\end{eqnarray}%
By (\ref{gianna}),

\begin{eqnarray}
\left\vert \Sigma _{\varepsilon ,t}\right\vert &\leq &\left\vert B\left( 
\hat{q}_{\varepsilon }(t),\eta ^{\frac{1}{8}}\left( 1+2\eta ^{\frac{1}{8}%
}\right) \right) \right\vert -\left\vert B\left( \hat{q}_{\varepsilon
}(t),\eta ^{\frac{1}{8}}\left( 1-2\eta ^{\frac{1}{8}}\right) \right)
\right\vert  \notag \\
&=&\omega _{N}\left[ \eta ^{\frac{1}{8}}\left( 1+2\eta ^{\frac{1}{8}}\right) %
\right] ^{N}-\omega _{N}\left[ \eta ^{\frac{1}{8}}\left( 1-2\eta ^{\frac{1}{8%
}}\right) \right] ^{N}  \notag \\
&=&\omega _{N}\eta ^{\frac{N}{8}}\left[ \left( 1+2\eta ^{\frac{1}{8}}\right)
^{N}-\left( 1-2\eta ^{\frac{1}{8}}\right) ^{N}\right]  \notag \\
&\leq &\omega _{N}\eta ^{\frac{N}{8}}\cdot 5N\eta ^{\frac{1}{8}}\leq
const.\eta ^{\frac{N+1}{8}}.  \label{ghi6}
\end{eqnarray}

So, by (\ref{ghi1}),....(\ref{ghi6})

\begin{eqnarray*}
\left\vert K_{\varepsilon }(t)\right\vert &\leq &\int_{\Sigma _{\varepsilon
,t}}\left\vert (x-q_{\varepsilon })\left[ u_{\varepsilon }^{2}\nabla
S_{\varepsilon }\cdot \nabla \rho _{\varepsilon }-\nabla \cdot J_{\rho
_{\varepsilon }}\right] \right\vert dx \\
&\leq &\underset{x\in \Sigma _{\varepsilon ,t}}{\sup }\left\vert
x-q_{\varepsilon }\right\vert \cdot \left[ \int_{\Sigma _{\varepsilon
,t}}\left\vert u_{\varepsilon }^{2}\nabla S_{\varepsilon }\cdot \nabla \rho
_{\varepsilon }\right\vert dx+\int_{\Sigma _{\varepsilon ,t}}\left\vert
\left( \nabla \cdot J_{\rho _{\varepsilon }}\right) \right\vert dx\right] \\
&\leq &\underset{x\in \Sigma _{\varepsilon ,t}}{\sup }\left\vert
x-q_{\varepsilon }\right\vert \cdot \left[ \underset{x\in \Sigma
_{\varepsilon ,t}}{\sup }\left\vert \nabla \rho _{\varepsilon }\right\vert
\cdot \int_{\Sigma _{\varepsilon ,t}}\left\vert u_{\varepsilon }^{2}\nabla
S_{\varepsilon }\right\vert +\underset{x\in \Sigma _{\varepsilon ,t}}{\sup }%
\left\vert \nabla \cdot J_{\rho _{\varepsilon }}\right\vert \cdot
\int_{\Sigma _{\varepsilon ,t}}dx\right] \\
&\leq &3\eta ^{1/8}\left[ const.\eta ^{-1/4}\cdot \eta ^{1/2}+const.\eta
^{-1/4}\cdot \left\vert \Sigma _{\varepsilon ,t}\right\vert \right] \\
&\leq &const.\eta ^{1/8}\left[ \eta ^{-1/4}\cdot \eta ^{1/2}+\eta
^{-1/4}\cdot \eta ^{\frac{N+1}{8}}\right] \leq const.\ \eta ^{1/8}.
\end{eqnarray*}%
Then, by Lemma \ref{lemmaconc}, 
\begin{equation}
\left\vert K_{\varepsilon }(t)\right\vert \rightarrow 0  \label{grulla}
\end{equation}%
uniformly in $t$.

Now, let us estimate $\left\vert H_{1,\varepsilon }(t)\right\vert ;$ by (\ref%
{ghi4}) and (\ref{ghi3}) we have%
\begin{eqnarray*}
\left\vert H_{1,\varepsilon }(t)\right\vert &\leq &\int_{\Sigma
_{\varepsilon ,t}}\left\vert u_{\varepsilon }^{2}\nabla S_{\varepsilon
}\left( \nabla \cdot J_{\rho _{\varepsilon }}\right) \right\vert dx \\
&\leq &\underset{x\in \Sigma _{\varepsilon ,t}}{\sup }\left\vert \nabla
\cdot J_{\rho _{\varepsilon }}\right\vert \cdot \int_{\Sigma _{\varepsilon
,t}}\left\vert u_{\varepsilon }^{2}\nabla S_{\varepsilon }\right\vert \\
&\leq &const.\frac{1}{\eta ^{1/4}}\cdot \eta ^{\frac{1}{2}}=const.\eta ^{1/4}
\end{eqnarray*}%
By the above estimate,%
\begin{equation}
\left\vert H_{1,\varepsilon }(t)\right\vert \rightarrow 0.  \label{grulla2}
\end{equation}

We recall that 
\begin{equation*}
\int |\Psi _{\varepsilon }|^{2}=1-o(1)
\end{equation*}
when $\varepsilon \rightarrow 0$, and that $\mathrm{supp}\Psi _{\varepsilon
}\subset B(\hat{q}_{\varepsilon },\ \eta ^{1/8}+2\eta ^{1/4})$. We have

\begin{equation*}
\nabla V(q_{\varepsilon }(t))=\left( 1+o(1)\right) \int \nabla
V(q_{\varepsilon }(t))\left\vert \Psi _{\varepsilon }\right\vert ^{2}
\end{equation*}%
and so 
\begin{eqnarray*}
|H_{2}(t)| &=&\left\vert \int \nabla V(x)\left\vert \Psi _{\varepsilon
}\right\vert ^{2}dx-\nabla V(q_{\varepsilon }(t))\right\vert \\
&=&\int |\nabla V(x)-\nabla V(q_{\varepsilon })|\left\vert \Psi
_{\varepsilon }\right\vert ^{2}dx+o(1)\int \nabla V(q_{\varepsilon
}(t))\left\vert \Psi _{\varepsilon }\right\vert ^{2} \\
&\leq &||V^{\prime \prime }||_{C^{0}(\mathbb{R}^{N})}\int \left\vert
x-q_{\varepsilon }\right\vert \left\vert \Psi _{\varepsilon }\right\vert
^{2}dx+o(1)\int \nabla V(q_{\varepsilon }(t))\left\vert \Psi _{\varepsilon
}\right\vert ^{2} \\
&\leq &o(1)\left( ||V^{\prime \prime }||_{C^{0}(\mathbb{R}^{N})}+||\nabla
V||_{C^{0}(\mathbb{R}^{N})}\right) =o(1)
\end{eqnarray*}%
for all $t$. By the above equation, (\ref{grulla}) and (\ref{grulla2}), the (%
\ref{qpuntostima}) follows.

Let $\mathbf{P=P}\left( \psi _{\varepsilon }\right) $ be defined by (\ref{P}%
). By the definitions of $p_{\varepsilon },\ $and (\ref{ghi2.5}), for every $%
t\in \mathbb{R}$, we have that 
\begin{eqnarray}
\left\vert p_{\varepsilon }(t)-\mathbf{P}(t)\right\vert &=&\left\vert \int
\nabla S\left( \left\vert \Psi _{\varepsilon }\right\vert
^{2}-u_{\varepsilon }^{2}\right) dx\right\vert  \notag \\
&\leq &\int_{\mathbb{R}^{N}\smallsetminus B(\hat{q}_{\varepsilon },\ \sqrt{%
\varepsilon })}\left\vert \nabla S\right\vert \ u_{\varepsilon }^{2}\
dx=c_{0}(\varepsilon )  \label{p-p}
\end{eqnarray}%
where $c_{1}(\varepsilon )\rightarrow 0$ as $\varepsilon \rightarrow 0.$ By (%
\ref{20}) 
\begin{equation*}
\mathbf{\dot{P}=}\int \left( -u_{\varepsilon }^{2}\nabla V+\nabla \cdot
T\right) dx
\end{equation*}%
and since $T\in L^{1}(\mathbb{R}^{N}),\ \mathbf{\dot{P}=}-\int
u_{\varepsilon }^{2}\nabla V\ dx.$ So, by (\ref{ppunto2}) and (\ref{effe})%
\begin{eqnarray*}
\dot{p}_{\varepsilon }-\mathbf{\dot{P}} &\mathbf{=}&\int \nabla V\left(
u_{\varepsilon }^{2}-\left\vert \Psi _{\varepsilon }\right\vert ^{2}\right)
dx-\int_{\Sigma _{\varepsilon ,t}}\left[ T\cdot \nabla \rho _{\varepsilon
}+u_{\varepsilon }^{2}\nabla S_{\varepsilon }\left( \nabla \cdot J_{\rho
_{\varepsilon }}\right) \right] dx \\
&=&\int \nabla V\left( u_{\varepsilon }^{2}-\left\vert \Psi _{\varepsilon
}\right\vert ^{2}\right) dx+F_{\varepsilon }(q_{\varepsilon })-\int_{\Sigma
_{\varepsilon ,t}}u_{\varepsilon }^{2}\nabla S_{\varepsilon }\left( \nabla
\cdot J_{\rho _{\varepsilon }}\right) dx.
\end{eqnarray*}%
Then, by (\ref{grulla2}) and Lemma \ref{lemmaconc}, 
\begin{eqnarray*}
\left\vert F_{\varepsilon }(q_{\varepsilon })-\left( \dot{p}_{\varepsilon }-%
\mathbf{\dot{P}}\right) \right\vert &=&\left\vert \int_{\Sigma _{\varepsilon
,t}}u_{\varepsilon }^{2}\nabla S_{\varepsilon }\left( \nabla \cdot J_{\rho
_{\varepsilon }}\right) dx-\int \nabla V\left( u_{\varepsilon
}^{2}-\left\vert \Psi _{\varepsilon }\right\vert ^{2}\right) dx\right\vert \\
&\leq &o(1)+\left\Vert \nabla V\right\Vert _{L^{\infty }}\int \left\vert
u_{\varepsilon }^{2}-\left\vert \Psi _{\varepsilon }\right\vert
^{2}\right\vert \\
&\leq &o(1)+\left\Vert \nabla V\right\Vert _{L^{\infty }}\int_{\mathbb{R}%
^{N}\smallsetminus B(\hat{q}_{\varepsilon },\ \sqrt{\varepsilon }%
)}u_{\varepsilon }^{2}\ dx=c_{1}(\varepsilon )
\end{eqnarray*}%
where $c_{1}(\varepsilon )\rightarrow 0$ as $\varepsilon \rightarrow 0.\ $%
Finally by (\ref{p-p}), $\forall \tau _{0},\tau _{1},\ $ 
\begin{eqnarray*}
\left\vert \int_{\tau _{0}}^{\tau _{1}}F_{\varepsilon }(q_{\varepsilon })\
dt\right\vert &=&\left\vert \int_{\tau _{0}}^{\tau _{1}}\left( \dot{p}%
_{\varepsilon }-\mathbf{\dot{P}}\right) \ dt\right\vert +c_{1}(\varepsilon
)\left( \tau _{1}-\tau _{0}\right) \\
&\leq &\left\vert p_{\varepsilon }(\tau _{1})-\mathbf{P}(\tau
_{1})\right\vert +\left\vert p_{\varepsilon }(\tau _{0})-\mathbf{P}(\tau
_{0})\right\vert +c_{1}(\varepsilon )\left( \tau _{1}-\tau _{0}\right) \\
&\leq &2c_{0}(\varepsilon )+c_{1}(\varepsilon )\left( \tau _{1}-\tau
_{0}\right) \leq c(\varepsilon )\left( 1+\left\vert \tau _{1}-\tau
_{0}\right\vert \right)
\end{eqnarray*}%
with a suitable choice of $c(\varepsilon ).$

$\square $

\bigskip

Collecting the previous results, we get our main theorem and Cor. \ref{cor1}%
:\bigskip

\textbf{Proof of Th. \ref{teo1}} By the def. (\ref{aa}),(\ref{bb}), Lemma %
\ref{lemmaconc} and Th. \ref{kappa}, Theorem \ref{teo1} holds with\textbf{\ }%
\begin{equation*}
R_{\varepsilon }=\eta ^{\frac{1}{8}}\left( 1+2\eta ^{\frac{1}{8}}\right) .
\end{equation*}

$\square $

\bigskip

\textbf{Proof of Cor. \ref{cor1}.} We rewrite (\ref{probcauchy}), (\ref{got}%
) and (\ref{got1}) in integral form and we get%
\begin{equation}
\left\{ 
\begin{array}{l}
q_{\varepsilon }(t)=q_{\varepsilon }(0)+\int_{0}^{t}\frac{p_{\varepsilon }(s)%
}{m_{\varepsilon }(s)}\ ds+\int_{0}^{t}K_{\varepsilon }(s)ds \\ 
p_{\varepsilon }(t)=p_{\varepsilon }(0)-\int_{0}^{t}\nabla V(q_{\varepsilon
}(s))ds+\int_{0}^{t}\left[ F_{\varepsilon }(q_{\varepsilon })+H_{\varepsilon
}(s)\right] ds%
\end{array}%
\right. 
\end{equation}%
\begin{equation}
\left\{ 
\begin{array}{l}
\mathfrak{q}(t)=q_{\varepsilon }(0)+\int_{0}^{t}\frac{\mathfrak{p}(s)}{m}\ ds
\\ 
\mathfrak{p}(t)=p_{\varepsilon }(0)-\int_{0}^{t}\nabla V(\mathfrak{q}(s))ds%
\end{array}%
\right. 
\end{equation}%
and hence, for any $|t|\leq T$%
\begin{eqnarray*}
\left\vert q_{\varepsilon }(t)-\mathfrak{q}(t)\right\vert  &\leq
&\int_{0}^{t}\left\vert \frac{p_{\varepsilon }(s)}{m_{\varepsilon }(s)}-%
\frac{\mathfrak{p}(s)}{m}\right\vert \ ds+\int_{0}^{t}\left\vert
K_{\varepsilon }(s)\right\vert ds \\
&\leq &L_{1}\int_{0}^{t}\left\vert p_{\varepsilon }(s)-\mathfrak{p}%
(s)\right\vert \ ds+\alpha _{1}(\varepsilon )
\end{eqnarray*}%
where, by (\ref{qpuntostima}), $\alpha _{1}(\varepsilon )\rightarrow 0$ as $%
\varepsilon \rightarrow 0$ and%
\begin{eqnarray*}
\left\vert p_{\varepsilon }(t)-\mathfrak{p}(t)\right\vert  &\leq
&\int_{0}^{t}\left\vert \nabla V(q_{\varepsilon }(s))-\nabla V(\mathfrak{q}%
(s))\right\vert ds+\left\vert \int_{0}^{t}F_{\varepsilon }(q_{\varepsilon
})ds\right\vert +\int_{0}^{t}\left\vert H_{\varepsilon }(s)\right\vert ds \\
&\leq &L_{2}\int_{0}^{t}\left\vert q_{\varepsilon }(s)-\mathfrak{q}%
(s)\right\vert \ ds+\alpha _{2}(\varepsilon )
\end{eqnarray*}%
where, by (\ref{qpuntostima}), $\alpha _{2}(\varepsilon )\rightarrow 0$ as $%
\varepsilon \rightarrow 0.$ Then, setting $z_{\varepsilon }(t)=\left\vert
q_{\varepsilon }(t)-\mathfrak{q}(t)\right\vert +\left\vert p_{\varepsilon
}(t)-\mathfrak{p}(t)\right\vert ,\ $we have 
\begin{equation*}
z_{\varepsilon }(t)\leq L\int_{0}^{t}z_{\varepsilon }(s)\ ds+\alpha
(\varepsilon )
\end{equation*}%
with a suitable choice of $L$ and $\alpha (\varepsilon ).$ Now, by the
Gronwall inequality, we have%
\begin{equation*}
z_{\varepsilon }(t)\leq \alpha (\varepsilon )e^{Lt}
\end{equation*}%
and from here, we get the conclusion.

$\square $

\end{document}